\documentclass[lettersize,journal]{IEEEtran}
\usepackage{amsmath,amsfonts}
\usepackage{algorithmic}
\usepackage{algorithm}
\usepackage{array}
\usepackage{booktabs}
\usepackage[caption=false,font=normalsize,labelfont=sf,textfont=sf]{subfig}
\usepackage{textcomp}
\usepackage{stfloats}
\usepackage{url}
\usepackage{verbatim}
\usepackage{graphicx}
\usepackage{cite}
\usepackage{orcidlink}

\begin{document}

\title{Information Security Risk Preparedness in Electronic Data Interchange: A Comparative Assessment of General IT and Healthcare Environments}

\author{%
Naresh Kshetri~\IEEEmembership{Senior Member, IEEE}~\orcidlink{0000-0002-3282-7331},
Bishwo Prakash Pokharel~\orcidlink{0000-0001-6516-2105},
Arati Bastakoti~\orcidlink{0009-0002-9583-4552},
and Anjana Bastakoti~\orcidlink{0009-0001-8088-6117}%
\thanks{Naresh Kshetri is with the Department of Cybersecurity, Rochester Institute of Technology, Rochester, New York 14623 USA, 
 e-mail: (naresh.kshetri@ieee.org). Bishwo Prakash Pokharel is with Department of Computer Science, University of Windsor, Ontario, Canada. Arati Bastakoti is with Nepal College of Information Technology, Nepal. Anjana Bastakoti is with Universal College of Medical Sciences, Bhairahawa, Nepal }
\thanks{Manuscript received mmm dd, yyyy; revised mmm dd, yyyy.}}


\maketitle

\begin{abstract}
Electronic Data Interchange (EDI) supports the exchange of structured information across business and healthcare environments, making information security and organizational risk management critical to reliable digital transactions. This study empirically assesses security-control preparedness and organizational risk-taking ability in EDI-related environments through a comparative analysis of General IT Professionals (GIT) and Healthcare EDI Professionals (HEP). A quantitative, cross-sectional survey was conducted with 100 respondents, equally divided between the two groups. Security Control Preparedness (SCP) and Threat Perception (TP) were validated using reliability and exploratory factor analyses, followed by group comparisons, correlation analysis, and multiple linear regression. HEP respondents reported significantly higher SCP than GIT respondents (\(M=4.70\) vs. \(3.89\), \(p<0.001\), \(d=1.06\)) and higher Organizational Risk-Taking Ability (ORTA) (\(M=6.14\) vs. \(5.28\), \(p=0.001\), \(d=0.66\)). In contrast, GIT respondents reported higher threat perception (\(M=2.95\) vs. \(2.04\), \(p<0.001\), \(d=0.72\)). SCP was positively associated with ORTA (\(r=0.479\), \(p<0.001\)), whereas TP showed a significant negative association (\(r=-0.610\), \(p<0.001\)). A parsimonious regression model incorporating SCP and TP explained 41.6\% of the variance in ORTA (\(R^2=0.416\), adjusted \(R^2=0.404\), \(p<0.001\)). Based on these empirical findings, an EDI Risk Management Framework (EDI-RMF) is proposed integrating governance, human and organizational security, technical safeguards, physical protection, and continuous monitoring. The findings provide empirical evidence for strengthening information-security risk management in EDI-enabled environments, particularly within developing-economy contexts.
\end{abstract}


\section{Introduction}

\IEEEPARstart{E}{lectronic} Data Interchange (EDI) enables standardized electronic transactions between organizations, reducing manual processing and supporting efficient information exchange. However, reliance on digitally exchanged information introduces security and organizational risks when threats exploit technical vulnerabilities, human error, or inadequate controls~\cite{ref1}. Protecting EDI therefore requires an integrated approach to information security and risk management.

\subsection{EDI and Information Security Context}

EDI security concerns the confidentiality, integrity, availability, and authenticity of exchanged information. These objectives protect sensitive business and healthcare data, preserve transaction accuracy, maintain service continuity, and verify message origins~\cite{ref2,ref3,ref4,ref5}. EDI supports transactions across healthcare, finance, manufacturing, logistics, and other sectors, improving processing efficiency while creating dependencies on reliable and secure information exchange~\cite{ref6}. In healthcare, for example, EDI supports administrative transactions such as claims processing, where security weaknesses may expose sensitive health information~\cite{ref7}.

Standards such as ANSI X12, UN/EDIFACT, and Health Level Seven (HL7) support structured commercial, international, and healthcare information exchange~\cite{ref8,ref9,ref10}. However, standardization alone does not ensure security. Effective protection also depends on governance, authentication, access control, data protection, user awareness, monitoring, incident response, and business continuity.

\subsection{EDI Security Challenges and Research Gap}

In Nepal, organizations adopting digital business processes must manage EDI-related risks involving commercial records, financial information, customer data, and other sensitive information. Inconsistent security procedures, limited formal training, and uneven implementation of technical and organizational controls may increase exposure to unauthorized access, data manipulation, information leakage, and service disruption. These challenges are particularly important when transactions involve external partners whose security practices also affect the reliability of information exchange.

Despite the importance of these issues, empirical evidence on EDI-related information-security preparedness in Nepal remains limited. In particular, further investigation is needed into differences in security-control implementation and threat perception across professional contexts and how these factors relate to organizational risk-related outcomes. This study addresses that gap by comparing General IT Professionals (GIT) and Healthcare EDI Professionals (HEP), examining Security Control Preparedness (SCP), Threat Perception (TP), and Organizational Risk-Taking Ability (ORTA), and using the findings to inform an EDI Risk Management Framework (EDI-RMF).

\subsection{Research Objectives and Questions}

This study has five objectives:

\begin{itemize}
    \item Assess Security Control Preparedness in EDI-related environments.
    \item Compare SCP, TP, and ORTA between GIT and HEP respondents.
    \item Identify group differences in individual security-awareness, technical, authentication, monitoring, continuity, and risk-management practices.
    \item Examine the associations of SCP and TP with ORTA.
    \item Develop an evidence-informed EDI Risk Management Framework integrating governance, human and organizational security, technical and physical safeguards, and continuous monitoring.
\end{itemize}

These objectives are addressed through five research questions:

\begin{itemize}
    \item \textbf{RQ1:} Does SCP differ significantly between GIT and HEP respondents?
    \item \textbf{RQ2:} Does reported TP differ significantly between the groups?
    \item \textbf{RQ3:} Which individual security-control and organizational practices differ between the groups?
    \item \textbf{RQ4:} How are SCP and TP associated with ORTA?
    \item \textbf{RQ5:} How can the identified security gaps inform an EDI-focused risk-management framework?
\end{itemize}

Four hypotheses guide the statistical analysis:

\begin{itemize}
    \item \textbf{H1:} SCP differs significantly between GIT and HEP respondents.
    \item \textbf{H2:} Reported TP differs significantly between GIT and HEP respondents.
    \item \textbf{H3:} Security-awareness and security-control practices differ significantly between the groups.
    \item \textbf{H4:} SCP and TP are significantly associated with ORTA.
\end{itemize}

\subsection{Research Contributions}

This study makes three principal contributions. First, it empirically compares reported information-security preparedness across general IT and healthcare EDI contexts. Second, it evaluates SCP and TP and examines their associations with ORTA, while identifying differences in specific security practices. Third, it synthesizes the findings into the proposed EDI Risk Management Framework (EDI-RMF), integrating governance, human and organizational security, technical safeguards, physical protection, and continuous monitoring. The framework offers a structured starting point for prioritizing security improvements in EDI-enabled environments, particularly in developing-economy contexts.

\section{Literature Review}

\subsection{EDI Information Security and Risk Management}

Electronic Data Interchange (EDI) enables organizations to exchange structured business information electronically across organizational and technological boundaries. Its security depends not only on the protection of individual systems but also on the integrity of exchange protocols, access controls, connected applications, and organizational processes. Hasan and Ali reviewed EDI representation, protocols, and security considerations, providing a technical foundation for understanding security challenges in EDI-enabled environments~\cite{HasanAli2019}. Similarly, Karata{\c{s}} and G{\"u}ltekin examined a secure design pattern for EDI-based data exchange in logistics and supply-chain settings, emphasizing the relevance of security-oriented design in interconnected information-exchange processes~\cite{KaratasGultekin2021}.

Beyond EDI-specific systems, research on data security and privacy provides additional context for managing risks across distributed information infrastructures. Xu \textit{et al.} examined information security and privacy challenges associated with big data and data-mining technologies~\cite{Xu2014}. Zhang \textit{et al.} surveyed security and privacy issues in edge-computing environments, while Yang \textit{et al.} reviewed data-security and privacy-protection mechanisms for cloud storage~\cite{Zhang2018,Yang2020}. Although these studies do not specifically evaluate the preparedness of EDI professionals, they illustrate the broader challenges of protecting information across interconnected systems and distributed data-processing environments.

EDI risk management consequently requires a coordinated approach that considers technical safeguards alongside organizational policies, accountability, and risk-assessment practices. Existing research provides useful perspectives on EDI security design and the protection of distributed information systems; however, these technical perspectives do not, by themselves, establish whether organizations consistently implement appropriate controls in practice. This distinction motivates an empirical assessment of reported security preparedness across different professional environments.

\subsection{IT Governance and Organizational Risk Management}

Information-security preparedness depends on the interaction of technical controls, organizational governance, staff responsibilities, and continuing risk-management activities. Established risk-management approaches emphasize identifying threats, evaluating their potential effects, selecting appropriate controls, and monitoring their effectiveness over time~\cite{ref11,ref12}. In EDI environments, these activities are particularly relevant because information exchange may involve multiple organizations, applications, and operational processes.

Risk assessment and security governance should therefore extend beyond the presence of individual technologies. Organizations also need procedures for reviewing vulnerabilities, managing access, maintaining audit records, responding to incidents, and recovering from operational disruptions. Research addressing security and privacy in big-data, edge-computing, and cloud-storage environments further illustrates the importance of considering data protection across interconnected infrastructure and service boundaries~\cite{Xu2014,Zhang2018,Yang2020}. Nevertheless, the implementation and effectiveness of such measures can vary with organizational resources, responsibilities, and risk exposure.

The present study examines organizational preparedness through reported security practices and perceptions rather than technical testing of deployed systems. This approach provides a basis for comparing the selected professional groups and identifying areas in which reported practices may require greater attention. It also supports the development of a proposed EDI Risk Management Framework (EDI-RMF) that connects organizational governance with operational and technical safeguards.

\subsection{Human Factors and Security Awareness}

Human factors are an important consideration in information-security risk management because the effectiveness of technical safeguards can be influenced by user behavior, staff awareness, and adherence to organizational procedures. Security-awareness activities, confidentiality obligations, authentication practices, and appropriate handling of sensitive information are therefore relevant components of organizational preparedness~\cite{ref13,ref14}. In EDI-related environments, personnel may interact with information systems and exchange processes that require consistent protection of confidential business or personal information.

The broader security literature also highlights the need to consider information protection across different computing environments rather than treating security as an isolated technical function~\cite{Xu2014,Zhang2018}. However, the existence of a policy or technical control does not necessarily demonstrate that it is consistently implemented or effective. Assessing reported practices alongside threat perceptions can therefore provide complementary evidence about organizational security preparedness.

In this study, security-control preparedness (SCP) represents reported implementation of selected security practices, whereas threat perception (TP) captures respondents' reported perception of specified threats. These constructs are analyzed separately because greater awareness of threats does not necessarily indicate stronger control implementation. Their relationships with organizational risk-taking ability (ORTA) are examined empirically rather than assumed in advance.

\subsection{Healthcare EDI Security as a Comparative Benchmark}

Healthcare information exchange presents important confidentiality and data-protection considerations because healthcare organizations handle sensitive information and depend on the availability and integrity of information systems. Research on healthcare industrial Internet of Things security has examined the use of blockchain in protecting healthcare-related data and connected systems, illustrating the broader interest in technical approaches to healthcare data protection~\cite{Khan2023}. Such work provides contextual support for considering healthcare information exchange within the wider information-security landscape, although healthcare IoT security is not equivalent to EDI security.

The healthcare group in the present study serves as a comparative professional context for examining reported security practices and perceptions. It is not assumed that healthcare EDI professionals necessarily possess stronger security expertise or operate more secure systems than general IT professionals. Instead, the comparison evaluates whether the two groups differ in their reported preparedness and identifies specific practices associated with those differences.

A related IEEE conference study, \textit{HR EDI Projects Risk Assessment}, addresses risk assessment in EDI-related projects and offers a project-risk perspective relevant to the broader discussion of EDI risk management~\cite{Livshitz2022}. However, project-level risk assessment and individual respondents' reported security practices represent different analytical levels. The present study focuses on the latter while considering how its findings may inform an organizational framework.

\subsection{Research Gap and Motivation for EDI-RMF}

The reviewed literature establishes the importance of secure EDI design, data protection, and risk management across interconnected information systems~\cite{HasanAli2019,KaratasGultekin2021,Xu2014,Zhang2018,Yang2020}. Related work also demonstrates the relevance of examining security risks in healthcare-related information environments and EDI projects~\cite{Khan2023,Livshitz2022}. However, these studies address different technical, contextual, or project-level questions and do not directly provide the same comparative evidence sought in the present research.

This study addresses that gap by comparing reported security-control preparedness, threat perception, and organizational risk-taking ability among General IT Professionals (GIT) and Healthcare EDI Professionals (HEP) in Nepal. It also examines differences in selected security practices and statistical associations among the study constructs. The results are used to inform the proposed EDI Risk Management Framework (EDI-RMF), which organizes relevant governance, organizational, technical, physical, and monitoring considerations into a structured approach to EDI risk management.

The framework is proposed on the basis of the reviewed literature and the study's empirical findings; it is not presented as an already validated intervention. Its effectiveness and applicability across different EDI environments require further evaluation.

\section{Research Methodology}

This study employed a quantitative, cross-sectional, comparative design to assess information-security preparedness in Electronic Data Interchange (EDI)-related environments. A structured questionnaire was used to compare General IT Professionals (GIT) and Healthcare EDI Professionals (HEP) and examine associations among Security Control Preparedness (SCP), Threat Perception (TP), and Organizational Risk-Taking Ability (ORTA). Descriptive and inferential analyses were conducted to evaluate these constructs and inform the proposed EDI Risk Management Framework (EDI-RMF)~\cite{ref18},~\cite{ref19}.

\subsection{Research Design and Sampling}

The study measured reported security practices and perceptions at a single point in time. The comparative design examined differences between GIT respondents working across Nepalese industries and HEP respondents involved in healthcare-related information exchange. The analysis also examined associations between security-related constructs and organizational risk-taking ability~\cite{ref19}.

Purposive, non-probability sampling was used to recruit respondents with relevant IT, information-security, or EDI-related experience~\cite{ref21}. The GIT group represented industries such as finance, manufacturing, logistics, and services, while the HEP group included professionals involved in health information exchange, medical billing, and clinical data interoperability. The final sample comprised 100 respondents, equally divided between GIT and HEP ($n=50$ per group). Because the sample was non-probabilistic, the findings should not be interpreted as representative of all Nepalese IT or healthcare EDI professionals~\cite{ref22}.

\subsection{Questionnaire and Measurement}

A structured questionnaire was developed using confidentiality, integrity, availability, and authenticity (CIAA) principles, relevant literature, and established information-security practices~\cite{ref23},~\cite{ref24}. It collected demographic and professional information and assessed organizational security controls, data protection, authentication, security awareness, threat perception, and risk-management practices. Security-control items covered backups, secure data disposal, access modification, confidentiality agreements, environmental safeguards, antivirus monitoring, intrusion detection or prevention, encryption, incident response, business continuity, risk assessment, audit logging, and removable-media practices.

Responses used predefined categorical and Likert-type scales documented in the study codebook. Negatively oriented five-point Likert items were reverse-scored as

\begin{equation}
X_{\mathrm{security}}=6-X_{\mathrm{raw}},
\end{equation}

so that higher scores represented stronger security practices or greater reported threat perception, as applicable. ORTA was measured using a single seven-point item ranging from extremely low to extremely high.

\subsection{Operationalization and Construct Validation}

SCP was calculated from eight security-practice items, TP from two threat-perception items, and ORTA from the single seven-point item. Internal consistency of the multi-item constructs was assessed using Cronbach's alpha, with values above 0.70 considered indicative of acceptable reliability~\cite{ref25}. Exploratory factor analysis examined the dimensionality of SCP, while the Kaiser--Meyer--Olkin (KMO) measure and Bartlett's test of sphericity assessed sampling adequacy and factorability. Internal-consistency analysis was not applicable to the single-item ORTA measure.

\subsection{Statistical Analysis}

Descriptive statistics summarized respondent characteristics and reported security practices. Welch's independent-samples $t$-test compared continuous construct scores between GIT and HEP respondents, with Mann--Whitney $U$ tests used as nonparametric sensitivity analyses. Pearson's chi-square tests examined group differences in categorical security practices, and Cram\'er's $V$ quantified their effect sizes. Effect sizes were considered alongside statistical significance.

Pearson's correlation analysis examined associations among SCP, TP, ORTA, and selected organizational security variables~\cite{ref26}. Multiple linear regression assessed the joint association of SCP and TP with ORTA~\cite{ref27}:

\begin{equation}
ORTA=\beta_{0}+\beta_{1}SCP+\beta_{2}TP+\epsilon,
\end{equation}

where $\beta_{0}$ is the intercept, $\beta_{1}$ and $\beta_{2}$ are the predictor coefficients, and $\epsilon$ is the residual error. Model fit was evaluated using $R^{2}$, adjusted $R^{2}$, and the overall $F$ statistic. Predictor effects were assessed using coefficients, standard errors, $t$ statistics, confidence intervals, and $p$ values. Variance inflation factors were examined for multicollinearity. All tests were two-sided, with statistical significance set at $\alpha=0.05$.

The analysis proceeded through construct validation, descriptive assessment, group comparisons, correlation analysis, and regression. These procedures addressed the hypotheses concerning group differences in SCP and TP, differences in individual security practices, and associations between SCP, TP, and ORTA. The resulting evidence informed the proposed EDI-RMF.

\section{Data Analysis and Results}

The analysis was conducted on 100 valid responses, equally divided between General IT Professionals (GIT) and Healthcare EDI Professionals (HEP). The results are presented in five stages: respondent characteristics, measurement validation, comparative security-control analysis, association analysis, and multivariable regression. This structure separates descriptive differences in reported practices from statistical tests of the principal study constructs.

\subsection{Respondent Characteristics}

The final sample comprised 100 respondents, with 50 GIT respondents (50\%) and 50 HEP respondents (50\%). The balanced group composition provides an equal basis for comparative analysis.

In terms of Internet experience, 30 respondents (30\%) reported less than five years of experience, 36 (36\%) reported between five and ten years, and 34 (34\%) reported more than ten years. Thus, 70\% of respondents reported more than five years of Internet experience.

Professional roles were dominated by programmers, software developers, and other IT professionals, accounting for 69 respondents (69\%). Managers and students each represented 11 respondents (11\%), followed by self-employed or freelance professionals (5\%), unemployed respondents (2\%), and one respondent each in non-employed and officer-level categories (1\% each).

The age distribution was concentrated in the 20-29-year range. Respondents aged 25-29 constituted 40\% of the sample, followed by those aged 20-24 (36\%), 30-34 (16\%), 35 years and above (7\%), and below 20 years (1\%). Educational attainment comprised bachelor's degree holders (40\%), master's degree or above (32\%), and intermediate/+2 education (28\%).

These characteristics indicate that the sample contains substantial representation from technically oriented respondents and respondents with several years of digital-system experience. However, the descriptive profile is used primarily to contextualize the subsequent analyses and is not treated as evidence of population-level representativeness.
\begin{table}[!t]
\caption{Respondent Characteristics ($N=100$)}
\label{tab:respondent_profile}
\centering
\footnotesize
\begin{tabular}{llr}
\hline
\textbf{Characteristic} & \textbf{Category} & \textbf{$n$ (\%)} \\
\hline
User group & GIT & 50 (50.0) \\
            & HEP & 50 (50.0) \\
\hline
Internet experience & $<5$ years & 30 (30.0) \\
                    & 5--10 years & 36 (36.0) \\
                    & $>10$ years & 34 (34.0) \\
\hline
Role & Programmer/developer/IT professional & 69 (69.0) \\
     & Manager & 11 (11.0) \\
     & Student & 11 (11.0) \\
     & Self-employed/freelance & 5 (5.0) \\
     & Unemployed & 2 (2.0) \\
     & Non-employed & 1 (1.0) \\
     & Officer & 1 (1.0) \\
\hline
Age & $<20$ & 1 (1.0) \\
    & 20--24 & 36 (36.0) \\
    & 25--29 & 40 (40.0) \\
    & 30--34 & 16 (16.0) \\
    & $\geq35$ & 7 (7.0) \\
\hline
Education & Intermediate/+2 & 28 (28.0) \\
          & Bachelor's & 40 (40.0) \\
          & Master's or above & 32 (32.0) \\
\hline
\end{tabular}
\end{table}

\subsection{Measurement Validation}

Before conducting comparative and multivariable analyses, the principal multi-item constructs were evaluated for internal consistency and dimensionality. Two constructs were analyzed as multi-item scales: Security Control Preparedness (SCP) and Threat Perception (TP). Organizational Risk-Taking Ability (ORTA) was measured using a single seven-point item and therefore was not subjected to internal-consistency analysis.

\subsubsection{Security Control Preparedness}

SCP was constructed from eight security-practice items covering regular data backup, secure data destruction, paper disposal, access modification following employee termination or transfer, non-disclosure/confidentiality practices, environmental controls, antivirus monitoring, and intrusion detection/prevention management. The negatively oriented five-point responses were reverse-scored so that higher values represented stronger reported security practices.

The SCP scale demonstrated high internal consistency, with Cronbach's alpha of $\alpha=0.926$. Corrected item-total correlations ranged from 0.596 to 0.849, indicating that all eight items contributed meaningfully to the composite construct. The Kaiser--Meyer--Olkin measure of sampling adequacy was 0.890, and Bartlett's test of sphericity was significant, $\chi^{2}(28)=578.50$, $p<0.001$, supporting factorability of the correlation matrix.

Exploratory factor analysis indicated a dominant first factor with an eigenvalue of 5.350. The remaining eigenvalues were below 1.0, and parallel analysis supported a one-factor interpretation. The corresponding standardized loadings ranged from 0.620 to 0.898, providing empirical support for treating the eight items as a single SCP construct.

The resulting SCP scores differed substantially between the two groups. GIT respondents had a mean SCP score of $M=3.89$ ($SD=0.86$), whereas HEP respondents had a mean of $M=4.70$ ($SD=0.66$). Welch's independent-samples $t$-test indicated a statistically significant difference, $t(91.92)=-5.301$, $p<0.001$. The corresponding Mann--Whitney $U$ test also indicated a significant group difference, $U=493$, $p<0.001$. The standardized mean difference was large ($d=1.06$).

These results indicate a substantial difference in reported security-control preparedness between the two respondent groups. The finding is descriptive of the sampled groups and does not establish that professional background alone causes the observed difference.
\begin{table}[!t]
\caption{Reliability and Construct-Validation Results}
\label{tab:validation}
\centering
\footnotesize
\begin{tabular}{lcc}
\hline
\textbf{Measure} & \textbf{SCP} & \textbf{TP} \\
\hline
Number of items & 8 & 2 \\
Cronbach's $\alpha$ & 0.926 & 0.791 \\
KMO & 0.890 & -- \\
Bartlett's $\chi^2$ & 578.50 & -- \\
Bartlett's df & 28 & -- \\
Bartlett's $p$ & $<0.001$ & -- \\
Inter-item $r$ & -- & 0.655 \\
Factor structure & One factor & Two-item scale \\
Loading range & 0.620--0.898 & -- \\
\hline
\end{tabular}
\end{table}

\subsubsection{Threat Perception}

Threat Perception (TP) was measured using two items addressing perceived threats from hackers or unauthorized persons and system failure. The five-point responses were reverse-scored so that higher scores represented greater reported threat perception.

The two-item scale demonstrated acceptable internal consistency, with Cronbach's alpha of $\alpha=0.791$. The two items were positively correlated, with Pearson's $r=0.655$, $p<0.001$, supporting their combination as a TP measure.

The group comparison produced a different pattern from SCP. GIT respondents reported a higher mean TP score ($M=2.95$, $SD=1.29$) than HEP respondents ($M=2.04$, $SD=1.24$). The between-group difference was statistically significant ($p<0.001$), with an effect size of approximately $d=0.72$.

This result should be interpreted as a difference in \emph{reported threat perception}, rather than as evidence that either group objectively faces greater or lesser cybersecurity risk. The finding also demonstrates that stronger reported security-control preparedness and higher reported threat perception do not necessarily occur in the same direction across the two groups.

\subsubsection{Organizational Risk-Taking Ability}

ORTA was measured using a seven-point item ranging from `Extremely Low'' to `Extremely High.'' The GIT group reported a mean ORTA score of $M=5.28$ ($SD=1.36$; median $=5$), while the HEP group reported $M=6.14$ ($SD=1.25$; median $=7$).

Welch's $t$-test indicated a statistically significant difference between the groups, $t(97.30)=-3.303$, $p=0.0013$. The Mann--Whitney $U$ test similarly indicated a significant difference, $U=772.5$, $p<0.001$. The corresponding standardized mean difference was approximately $d=0.66$.

Because ORTA is based on a single self-reported item, this result should be interpreted as a difference in reported organizational risk-taking ability rather than as an independently verified organizational risk-capacity measure.

\subsection{Comparative Security-Control Practices}

Beyond the composite constructs, individual security practices were compared between GIT and HEP respondents. These analyses identify specific controls contributing to the observed differences in security preparedness.

\subsubsection{Policy, Governance, and Organizational Controls}

The reported implementation of selected organizational security controls was higher among HEP respondents for most measured practices. Data-protection policy implementation was reported by 58\% of GIT respondents and 74\% of HEP respondents. Incident-response and business-continuity practices were reported by 32\% and 74\%, respectively. Background checks were reported by 58\% of GIT respondents and 76\% of HEP respondents, while audit logging was reported by 36\% and 66\%, respectively.

Three categories—non-disclosure/confidentiality agreements, environmental controls, and formally documented access-control practices—showed no reported implementation in the corresponding survey items. These results indicate that substantial control gaps remained in both groups, even where other security practices were more frequently reported.

Across the measured policy and organizational controls, the average reported implementation rate was 35.4\% for GIT respondents and 62.9\% for HEP respondents. Pearson's chi-square tests identified statistically significant group differences for incident-response/business-continuity practices ($\chi^2=18.04$, $p<0.001$, Cram'er's $V=0.425$) and audit logging ($\chi^2=11.64$, $p=0.003$, $V=0.341$). The difference for data-protection policy implementation was not statistically significant ($p=0.129$), and the difference for background checks was also not statistically significant ($p=0.142$).

\subsubsection{Authentication, Endpoint, and Data-Protection Controls}

Differences were also observed in authentication and endpoint-protection practices. Screen-lock implementation was reported by 86\% of GIT respondents and 98\% of HEP respondents, although the group difference did not reach the 0.05 significance threshold ($p=0.065$). Complex password requirements were reported by 84\% of GIT respondents and 100\% of HEP respondents, with a significant group difference ($\chi^2=6.66$, $p=0.010$, $V=0.258$).

Hard-drive encryption showed one of the largest differences: 32\% of GIT respondents reported encryption compared with 84\% of HEP respondents. This difference was statistically significant, $\chi^2=27.80$, $p<0.001$, with Cram'er's $V=0.527$.

For mobile-device authentication or password protection, the reported implementation rates were 63.4\% for GIT respondents and 77.8\% for HEP respondents. The group difference was statistically significant, $\chi^2=22.87$, $p<0.001$, $V=0.478$.

Remote-access practices also differed between the groups. Among GIT respondents, 36\% reported using two-factor authentication through a secondary device, 34\% relied on passwords alone, and 30\% indicated that remote access was not applicable. Among HEP respondents, 24\% reported two-factor authentication, 12\% reported password-only authentication, and 64\% indicated that remote access was not applicable. The higher ``not applicable'' response among HEP respondents should be considered when interpreting the apparent differences in authentication practice.

\subsubsection{Security Awareness, Email, and Removable Media}

Security-awareness and endpoint practices showed mixed results. Formal training for identifying unusual or suspicious email was reported by 32\% of GIT respondents and 34\% of HEP respondents. Despite the similar percentages, the overall response distributions differed significantly, $\chi^2=11.19$, $p=0.004$, with Cram'er's $V=0.335$.

Removable-media control software was reported by 28\% of GIT respondents and 48\% of HEP respondents. The group difference was statistically significant, $\chi^2=10.70$, $p=0.005$, $V=0.327$.

Antivirus signatures were reported as current by 76\% of GIT respondents and 92\% of HEP respondents. Although the HEP percentage was higher, the group difference did not reach statistical significance at the 0.05 level ($p=0.089$).

Taken together, these results indicate that group differences were not uniform across all controls. Some controls, particularly encryption, business continuity, incident response, risk assessment, authentication, and audit logging, showed statistically meaningful differences, whereas other baseline controls did not.
\begin{figure}[!t]
\centering
\includegraphics[width=3.5 in]{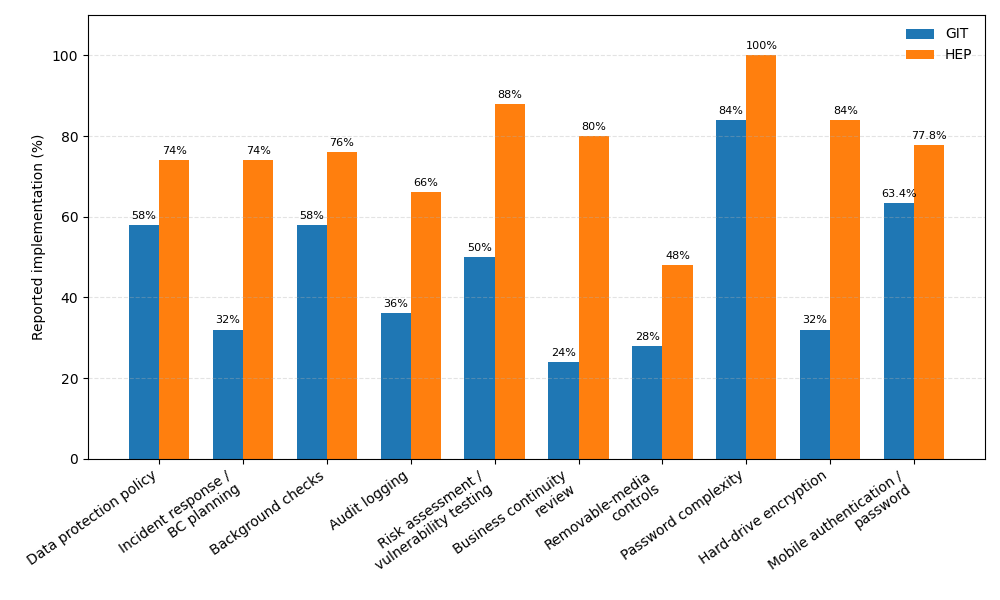}
\caption{Reported implementation of selected information-security controls among General IT Professionals (GIT) and Healthcare EDI Professionals (HEP).}
\label{fig:security_control_gap}
\end{figure}

\subsection{Organizational Risk-Management Practices}

Three organizational practices were particularly relevant to risk preparedness: risk assessment and vulnerability testing, business-continuity review, and incident-response planning. Risk assessment or vulnerability testing was reported by 50\% of GIT respondents and 88\% of HEP respondents, corresponding to a significant group difference of $\chi^2=16.90$, $p<0.001$, and $V=0.411$.

Business-continuity review showed an even larger difference, with reported implementation of 24\% among GIT respondents and 80\% among HEP respondents ($\chi^2=31.43$, $p<0.001$, $V=0.561$). Incident-response planning was reported by 32\% of GIT respondents and 74\% of HEP respondents ($\chi^2=18.04$, $p<0.001$, $V=0.425$).

The average implementation rate across these organizational risk-management indicators was 45.5\% for GIT respondents and 83.5\% for HEP respondents. These descriptive differences provide specific evidence of variation in reported organizational risk-management practices and inform the control priorities incorporated into the proposed EDI-RMF.
\begin{figure}[!t]
\centering
\includegraphics[width=3.5 in]{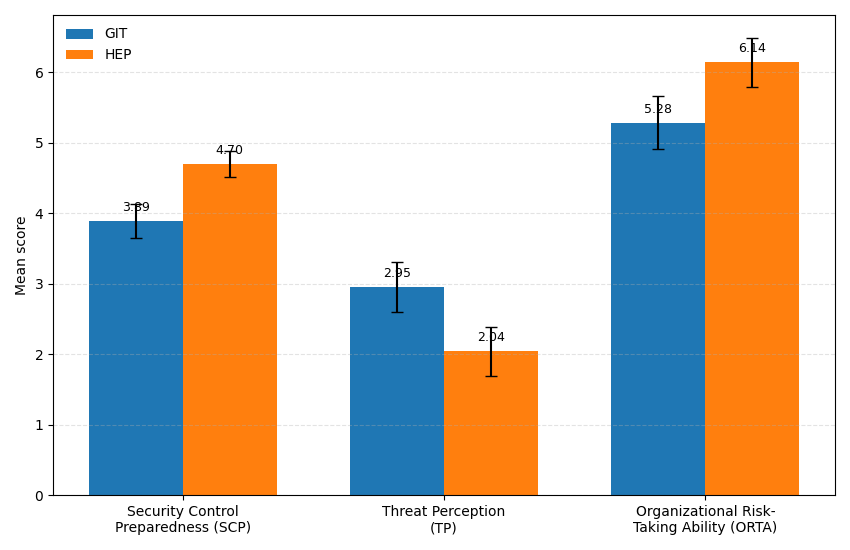}
\caption{Comparison of Security Control Preparedness (SCP), Threat Perception (TP), and Organizational Risk-Taking Ability (ORTA) between General IT Professionals (GIT) and Healthcare EDI Professionals (HEP). Error bars represent approximate 95\% confidence intervals.}
\label{fig:construct_comparison}
\end{figure}

\begin{table*}[!t]
\caption{Comparative Security-Control and Organizational Practice Results}
\label{tab:control_comparison}
\centering
\footnotesize
\begin{tabular}{lrrrrr}
\hline
\textbf{Security practice} &
\textbf{GIT (\%)} &
\textbf{HEP (\%)} &
\textbf{$\chi^2$} &
\textbf{$p$} &
\textbf{Cram\'er's $V$} \\
\hline
Data protection policy & 58.0 & 74.0 & -- & 0.129 & -- \\
Incident response / BC & 32.0 & 74.0 & 18.04 & $<0.001$ & 0.425 \\
Background checks & 58.0 & 76.0 & -- & 0.142 & -- \\
Audit logging & 36.0 & 66.0 & 11.64 & 0.003 & 0.341 \\
Risk assessment / vulnerability testing & 50.0 & 88.0 & 16.90 & $<0.001$ & 0.411 \\
Business continuity review & 24.0 & 80.0 & 31.43 & $<0.001$ & 0.561 \\
Email/phishing training & 32.0 & 34.0 & 11.19 & 0.004 & 0.335 \\
Removable-media controls & 28.0 & 48.0 & 10.70 & 0.005 & 0.327 \\
Password complexity & 84.0 & 100.0 & 6.66 & 0.010 & 0.258 \\
Hard-drive encryption & 32.0 & 84.0 & 27.80 & $<0.001$ & 0.527 \\
Mobile authentication/password & 63.4 & 77.8 & 22.87 & $<0.001$ & 0.478 \\
Screen lock & 86.0 & 98.0 & -- & 0.065 & -- \\
Antivirus currency & 76.0 & 92.0 & -- & 0.089 & -- \\
\hline
\end{tabular}
\end{table*}

\subsection{Association Between Security Constructs and Organizational Risk-Taking Ability}

Pearson's correlation analysis was conducted to examine relationships between the principal constructs. SCP was positively associated with ORTA ($r=0.479$, $p<0.001$), indicating that respondents reporting stronger security-control practices also tended to report higher organizational risk-taking ability.
\begin{table*}[!t]
\caption{Principal Construct Comparisons, Correlations, and Regression Results}
\label{tab:inferential_results}
\centering
\footnotesize
\begin{tabular}{llrrrrr}
\hline
\textbf{Analysis} &
\textbf{Variable} &
\textbf{GIT} &
\textbf{HEP} &
\textbf{Statistic} &
\textbf{$p$} &
\textbf{Effect/Estimate} \\
\hline
Group comparison & SCP & 3.89 $\pm$ 0.86 & 4.70 $\pm$ 0.66 &
$t(91.92)=-5.301$ & $<0.001$ & $d=1.06$ \\
Group comparison & TP & 2.95 $\pm$ 1.29 & 2.04 $\pm$ 1.24 &
-- & $<0.001$ & $d\approx0.72$ \\
Group comparison & ORTA & 5.28 $\pm$ 1.36 & 6.14 $\pm$ 1.25 &
$t(97.30)=-3.303$ & 0.0013 & $d=0.66$ \\
\hline
Correlation & SCP--ORTA & \multicolumn{2}{c}{--} &
$r=0.479$ & $<0.001$ & Positive \\
Correlation & TP--ORTA & \multicolumn{2}{c}{--} &
$r=-0.610$ & $<0.001$ & Negative \\
Correlation & Risk assessment--ORTA & \multicolumn{2}{c}{--} &
$r=0.176$ & 0.080 & -- \\
Correlation & Business continuity--ORTA & \multicolumn{2}{c}{--} &
$r=0.296$ & 0.003 & Positive \\
Correlation & Incident response/BC--ORTA & \multicolumn{2}{c}{--} &
$r=0.039$ & 0.698 & -- \\
\hline
Regression & SCP & \multicolumn{2}{c}{--} &
$B=0.379$ & 0.008 & 95\% CI [0.101,0.657] \\
Regression & TP & \multicolumn{2}{c}{--} &
$B=-0.505$ & $<0.001$ & 95\% CI [-0.684,-0.325] \\
\hline
\multicolumn{2}{l}{Regression model} &
\multicolumn{2}{c}{$R^2=0.416$} &
$F(2,97)=34.55$ &
$<0.001$ &
Adj. $R^2=0.404$ \\
\hline
\end{tabular}
\end{table*}

\begin{figure}[!t]
\centering
\includegraphics[width=3.5 in]{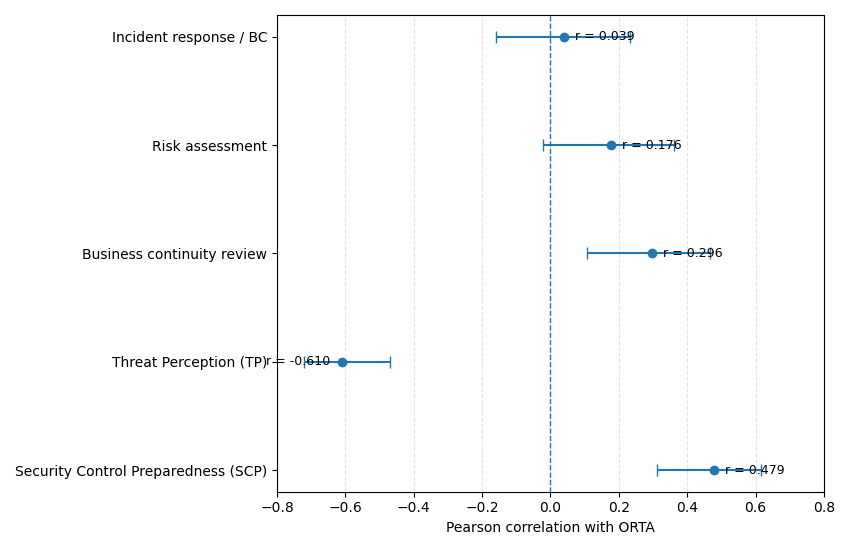}
\caption{Associations between principal security and organizational variables and Organizational Risk-Taking Ability (ORTA). Error bars represent approximate 95\% confidence intervals for Pearson correlation coefficients.}
\label{fig:orta_correlations}
\end{figure}

TP demonstrated a significant negative association with ORTA ($r=-0.610$, $p<0.001$). Thus, within the sampled respondents, higher reported threat perception was associated with lower reported organizational risk-taking ability. This association should not be interpreted causally because the cross-sectional design does not establish temporal direction.

Among selected individual organizational practices, business-continuity review was positively associated with ORTA ($r=0.296$, $p=0.003$), whereas risk assessment implementation did not reach statistical significance ($r=0.176$, $p=0.080$). Incident-response planning showed essentially no linear association with ORTA ($r=0.039$, $p=0.698$). These results further support the use of construct-level analysis rather than relying exclusively on individual control indicators.

\subsection{Multiple Linear Regression}

A parsimonious multiple linear regression model was estimated to examine the joint association of SCP and TP with ORTA:

\begin{equation}
ORTA=\beta_{0}+\beta_{1}SCP+\beta_{2}TP+\epsilon.
\end{equation}

The model was statistically significant, $F(2,97)=34.55$, $p<0.001$, and explained 41.6\% of the variance in ORTA ($R^{2}=0.416$; adjusted $R^{2}=0.404$).

Both predictors made statistically significant contributions to the model. SCP was positively associated with ORTA ($B=0.379$, $SE=0.140$, $t=2.709$, $p=0.008$, 95\% CI $[0.101,0.657]$). TP was negatively associated with ORTA ($B=-0.505$, $SE=0.091$, $t=-5.571$, $p<0.001$, 95\% CI $[-0.684,-0.325]$). Variance inflation factors were approximately 3.06 for both predictors, indicating that the model did not exhibit severe multicollinearity.

The regression results complement the correlation analysis. Higher SCP was associated with higher reported ORTA, whereas higher TP was associated with lower reported ORTA after accounting for SCP. These coefficients represent statistical associations within the sampled population and should not be interpreted as causal effects.

\subsection{Hypothesis-Related Findings}

The hypothesis tests synthesize the comparative and associational results reported above. H1 is supported: Security Control Preparedness (SCP) differed significantly between General IT Professionals (GIT) and Healthcare EDI Professionals (HEP), with higher SCP among HEP respondents ($M=4.70$, $SD=0.66$ vs. $M=3.89$, $SD=0.86$; Welch $t(91.92)=-5.301$, $p<0.001$). H2 is also supported: reported Threat Perception (TP) differed significantly between the groups, with higher scores among GIT respondents ($M=2.95$, $SD=1.29$ vs. $M=2.04$, $SD=1.24$; $p<0.001$). H3 is supported for selected security-awareness and organizational practices, with significant group differences observed in phishing training, removable-media controls, password complexity, encryption, authentication, audit logging, risk assessment, business continuity, and incident-response practices. H4 is supported by the association analysis: SCP was positively correlated with Organizational Risk-Taking Ability (ORTA) ($r=0.479$, $p<0.001$), whereas TP was negatively correlated with ORTA ($r=-0.610$, $p<0.001$). The parsimonious regression model containing SCP and TP explained 41.6\% of the variance in ORTA ($R^2=0.416$, adjusted $R^2=0.404$; $F(2,97)=34.55$, $p<0.001$).

\begin{table}[!t]
\centering
\caption{Summary of Hypothesis Testing}
\label{tab:hypothesis_summary}
\renewcommand{\arraystretch}{1.05}
\begin{tabular}{p{0.7cm}p{1.0cm}p{4.9cm}}
\toprule
\textbf{Hyp.} & \textbf{Decision} & \textbf{Primary Evidence} \\
\midrule
H1 & Supported &
SCP differed significantly between GIT and HEP ($M=3.89$ vs. $4.70$; $p<0.001$). \\

H2 & Supported &
TP differed significantly between GIT and HEP ($M=2.95$ vs. $2.04$; $p<0.001$). \\

H3 & Supported &
Significant differences occurred across selected security-awareness and control practices, including phishing training, encryption, authentication, risk assessment, business continuity, and incident response. \\

H4 & Supported &
SCP correlated positively with ORTA ($r=0.479$), TP negatively with ORTA ($r=-0.610$), and the regression model explained 41.6\% of ORTA variance ($p<0.001$). \\
\bottomrule
\end{tabular}
\end{table}

Overall, the results provide convergent evidence of group differences in reported security preparedness and threat perception and demonstrate significant associations between security-related constructs and organizational risk-taking ability. These findings provide the empirical basis for the EDI Risk Management Framework (EDI-RMF), which translates the identified gaps into governance, human and organizational, technical, physical, and continuous-monitoring dimensions.
 
\section{Discussion}

The empirical findings indicate meaningful differences in reported information-security preparedness between General IT Professionals (GIT) and Healthcare EDI Professionals (HEP) across several security and organizational dimensions. Rather than interpreting these differences as evidence that one professional group is inherently more security-capable than the other, the results suggest that the organizational and regulatory context in which EDI-related activities are performed may be associated with differences in security-control implementation, threat perception, and organizational risk-taking ability. This interpretation is consistent with established information-security governance literature emphasizing the importance of organizational policies, accountability, risk management, and institutional controls in addition to technical safeguards~\cite{ref28}.

\subsection{Security-Control Preparedness and Organizational Context}

The strongest group-level difference was observed in the Security Control Preparedness (SCP) construct. HEP respondents reported significantly higher SCP than GIT respondents, and the difference remained significant under both parametric and non-parametric group-comparison tests. The individual-control analysis provides additional context for this finding: substantial differences were observed in risk assessment, business continuity review, incident response and continuity practices, audit logging, hard-drive encryption, removable-media controls, and mobile authentication.

These findings suggest that differences between the two groups extend beyond basic endpoint hygiene to organizational and preventive controls. Screen locking, password complexity, and antivirus currency were comparatively common across both groups, whereas controls associated with formal risk management, resilience, information protection, and organizational accountability showed larger differences. This pattern is consistent with information-security maturity perspectives in which organizational preparedness involves not only technical protection but also governance, risk assessment, continuity, monitoring, and recovery capabilities~\cite{ref29},~\cite{ref30}.

The findings should nevertheless be interpreted cautiously. The study does not establish that healthcare EDI specialization itself causes stronger security controls. Healthcare EDI environments may differ from general IT environments in regulatory requirements, data sensitivity, organizational structure, operational processes, and resource allocation. These contextual factors provide plausible explanations for the observed differences but were not independently measured in the present survey.

\subsection{Healthcare EDI as a Comparative Security Context}

Healthcare EDI provides an important comparative context because healthcare information systems commonly operate under formal requirements for protecting sensitive information and maintaining appropriate administrative, physical, and technical safeguards. Such environments can encourage greater institutionalization of security policies, accountability mechanisms, and risk-management processes~\cite{ref31},~\cite{ref32}.

The present findings are consistent with this contextual interpretation. HEP respondents reported higher implementation of several organizational and technical safeguards, particularly risk assessment, business continuity review, incident response, and encryption. However, the results should not be interpreted as demonstrating that all healthcare EDI organizations possess higher security maturity. The sample measures respondents' reported practices within the surveyed environments and does not constitute an independent audit of organizational compliance or security maturity.

This distinction is important for interpreting the proposed EDI Risk Management Framework (EDI-RMF). The purpose of using healthcare EDI as a comparative context is not to prescribe healthcare controls for all EDI environments, but to identify organizational practices that may be transferable to other EDI-enabled sectors when appropriately adapted to their risk profiles.

\subsection{Human Factors and Security Awareness}

The results also demonstrate that technical control implementation does not necessarily imply comprehensive security awareness. Although several technical and organizational controls differed significantly between the groups, formal email and phishing-related training remained comparatively limited. The observed difference in reported phishing-training practice was statistically significant, but the implementation rates were relatively low in both groups.

This finding reinforces the importance of treating human factors as an integral component of EDI security. EDI environments involve users who interact with credentials, business records, sensitive information, email communications, removable media, and administrative systems. Consequently, security preparedness cannot be represented adequately by infrastructure controls alone. Security awareness, training, accountability, and appropriate user behavior must operate together with technical safeguards~\cite{ref33}.

The construct-level analysis further showed that Threat Perception (TP) was significantly associated with Organizational Risk-Taking Ability (ORTA). The relationship was negative in the present sample. This finding should not be interpreted as evidence that greater threat awareness necessarily reduces organizational capability. Rather, respondents reporting greater perceived threat also tended to report lower organizational risk-taking ability. This association may reflect a more cautious perception of organizational risk, although the cross-sectional design does not permit determination of the underlying mechanism.

\subsection{Relationship Between Security Preparedness and Risk-Taking Ability}

An important contribution of the study is the examination of how security-related constructs relate to Organizational Risk-Taking Ability. SCP was positively associated with ORTA ($r=0.479$, $p<0.001$), whereas TP was negatively associated with ORTA ($r=-0.610$, $p<0.001$). The multiple regression model incorporating SCP and TP explained 41.6\% of the observed variance in ORTA.

The positive SCP--ORTA relationship suggests that organizations reporting stronger security-control preparedness also tend to report greater organizational capacity for risk-taking. One possible interpretation is that formal security controls may provide organizations with greater visibility, accountability, and control over information-security risks. However, this interpretation remains an association rather than a causal explanation.

The negative TP--ORTA relationship is particularly important because it illustrates that security preparedness and perceived threat are not interchangeable constructs. An organization may recognize substantial threats while simultaneously reporting lower confidence in its ability to undertake organizational risk. Conversely, stronger control preparedness may coexist with greater perceived organizational capacity to manage risk. These findings support the inclusion of both preventive controls and risk-perception considerations in an EDI security framework.

The regression results also provide a more parsimonious explanation than analyses based on individual technical controls. Rather than interpreting individual safeguards as independent drivers of organizational risk-taking, the final model considers the two validated constructs that capture broader dimensions of security preparedness and threat perception. This reduces redundancy among overlapping security-control variables and provides a clearer empirical basis for the proposed framework.

\subsection{Implications for Nepalese EDI Environments}

From a Nepalese EDI perspective, the findings identify several areas where organizations can strengthen information-security preparedness. The largest observed differences were associated with risk assessment, business continuity, incident response, encryption, audit logging, and authentication-related controls. These areas are particularly relevant to EDI because disruption, unauthorized disclosure, or manipulation of exchanged business information can affect not only an individual organization but also interconnected trading partners and service providers.

The findings therefore support a shift from predominantly technology-centered protection toward a more integrated risk-management approach. For organizations participating in EDI-enabled supply chains, security preparedness should encompass governance and policy, user awareness, technical safeguards, physical protection, incident response, continuity planning, and continuous monitoring. Such an approach is consistent with established information-security risk-management principles and provides a practical basis for adapting security controls to organizational context~\cite{ref28},~\cite{ref29}.

The study does not establish that Nepalese organizations face a uniform level of security deficiency, nor does the sample permit generalization to every Nepalese B2B organization. Instead, the findings provide empirical evidence of specific differences and control gaps within the surveyed population. These observations can inform organizations and policymakers when prioritizing security assessments and capacity-building initiatives.

\subsection{Implications for the EDI Risk Management Framework}

The empirical findings provide the rationale for the five dimensions of the proposed EDI-RMF. First, the differences in policy implementation, accountability, audit logging, and risk assessment support a \emph{Governance and Policy} dimension. Second, the observed limitations in security-awareness and phishing-related training support a \emph{Human and Organizational Security} dimension. Third, differences in encryption, authentication, endpoint protection, and related controls motivate the \emph{Technical and System Security} dimension. Fourth, organizational protection requirements support explicit consideration of \emph{Physical and Environmental Security}. Finally, the observed importance of risk assessment, incident response, business continuity, and monitoring supports a \emph{Continuous Monitoring and Improvement} dimension.

Accordingly, the EDI-RMF is not proposed solely as a theoretical taxonomy of security controls. It is derived from the empirical patterns observed in the present study and organizes those findings into a structured risk-management model. The framework is intended to help EDI-enabled organizations identify control gaps, establish governance responsibilities, strengthen human and technical safeguards, and continuously reassess their security posture.

\subsection{Overall Interpretation}

Taken together, the findings indicate that EDI information security is multidimensional and cannot be adequately represented by individual technical controls. Security Control Preparedness, Threat Perception, organizational practices, and risk-taking ability exhibit distinct but related patterns across the two respondent groups. The evidence particularly highlights the importance of organizational risk assessment, continuity and incident-response capabilities, encryption, authentication, security awareness, and governance.

The findings therefore provide an empirical foundation for the EDI-RMF proposed in the following section. Rather than assuming that a particular professional background determines security maturity, the framework emphasizes the organizational capabilities that can be assessed, implemented, monitored, and improved across different EDI environments.
\section{Proposed EDI Risk Management Framework}

The empirical findings identify differences in security-control preparedness, threat perception, organizational practices, and risk-taking ability across the two respondent groups. These findings indicate that EDI security cannot be addressed through isolated technical safeguards alone. Instead, effective EDI risk management requires coordinated governance, human and organizational controls, technical protection, physical safeguards, and continuous monitoring.

Based on these findings, this study proposes the \textit{EDI Risk Management Framework (EDI-RMF)} as a structured model for organizing and continuously improving information-security controls in EDI-enabled environments. The framework translates the empirical control gaps identified in Section IV and the implications discussed in Section V into five complementary security dimensions: 1) governance and policy, 2) human and organizational security, 3) technical and system security, 4) physical and environmental security, and 5) continuous monitoring and improvement.

Rather than treating these dimensions as independent or strictly sequential layers, the EDI-RMF adopts a cyclic and feedback-driven architecture. Governance establishes organizational direction and accountability; human and organizational controls translate policies into expected behavior; technical and physical safeguards implement protection mechanisms; and continuous monitoring provides evidence for reassessment and improvement. Feedback from incidents, audits, control assessments, and emerging threats is subsequently incorporated into governance and risk-management decisions.

\begin{figure}[!t]
\centering
\includegraphics[width=3.5in]{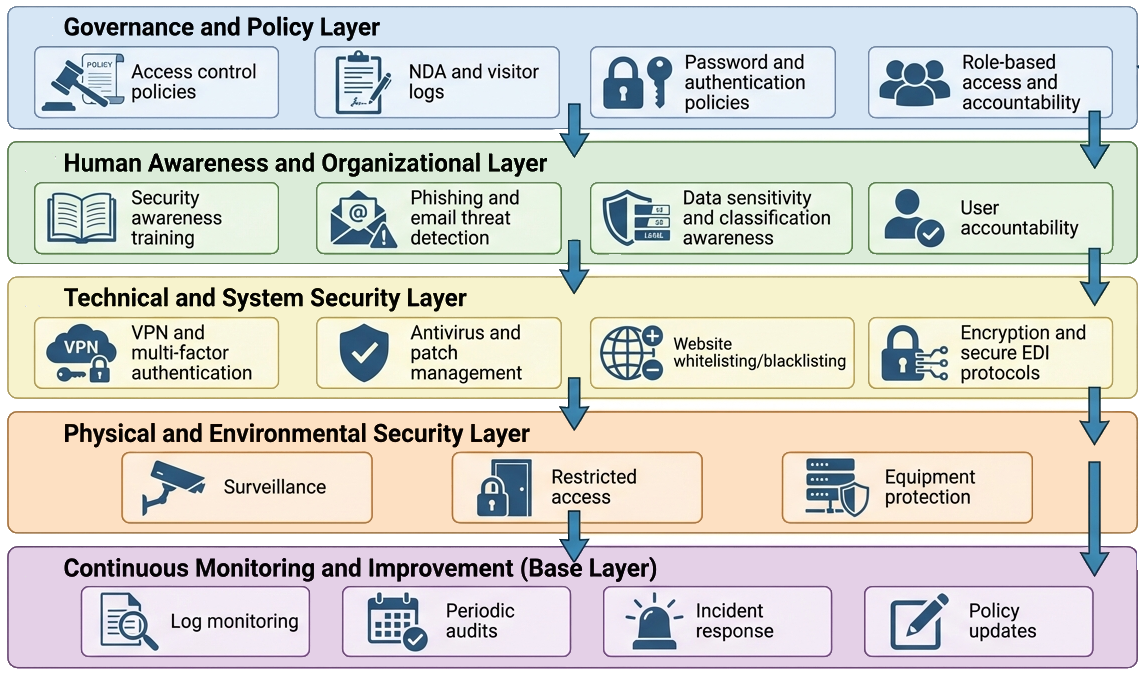}
\caption{Proposed EDI Risk Management Framework (EDI-RMF).}
\label{fig:edi-rmf}
\end{figure}

\subsection{Framework Rationale and Design Principles}

The EDI-RMF is guided by four principles. First, \textit{risk-based prioritization} requires organizations to identify and prioritize controls according to the sensitivity of exchanged information, business dependencies, threat exposure, and potential operational impact. Second, \textit{defense in depth} recognizes that no individual control is sufficient to protect interconnected EDI environments. Third, \textit{accountability} assigns responsibilities for security decisions, implementation, monitoring, and incident response. Fourth, \textit{continuous improvement} ensures that security controls are periodically reassessed as organizational conditions, technologies, threats, and regulatory requirements change.

The framework is particularly informed by the empirical gaps observed in risk assessment, business continuity, incident response, encryption, authentication, audit logging, removable-media controls, and security awareness. These findings provide the primary empirical motivation for the framework dimensions.

\subsection{Governance and Policy}

The Governance and Policy dimension establishes the strategic and organizational foundation of EDI risk management. It defines security responsibilities, control objectives, access boundaries, accountability mechanisms, and organizational requirements for protecting EDI information and services.

Core controls include information-security policies, data-protection requirements, access-control policies, role-based access control, authentication requirements, non-disclosure agreements, visitor management, segregation of duties, and defined security responsibilities. Governance should also establish requirements for periodic risk assessment, business continuity, incident response, audit logging, and third-party or trading-partner security.

The empirical findings provide direct support for emphasizing this dimension. Significant differences were observed in risk assessment, audit logging, business continuity review, and incident-response-related practices. The framework therefore positions governance not merely as documentation but as the mechanism through which security responsibilities, risk priorities, and control requirements are established and evaluated.

\subsection{Human and Organizational Security}

The Human and Organizational Security dimension addresses the behavioral and organizational factors that influence EDI risk. Policies are effective only when users understand their responsibilities and apply appropriate security practices during routine operations.

This dimension includes security-awareness training, phishing and email-threat recognition, data sensitivity and classification, credential protection, removable-media practices, user accountability, and periodic security training. Training should be role-specific and reinforced through periodic assessment rather than treated as a one-time compliance activity.

The survey findings indicate that human and organizational practices remain important areas for improvement. Although several technical controls were relatively common, reported formal email and phishing-related training remained limited. The framework therefore treats user awareness and organizational accountability as complementary to technical safeguards rather than as separate awareness activities.

\subsection{Technical and System Security}

The Technical and System Security dimension translates organizational requirements into enforceable technical safeguards for EDI systems, endpoints, networks, and exchanged information.

Key controls include strong authentication and multi-factor authentication, role-based access control, encryption for data at rest and in transit, secure EDI communication protocols, endpoint protection, vulnerability and patch management, secure configuration, network segmentation where appropriate, and security logging. VPNs and other secure connectivity mechanisms may be applied where remote access creates additional exposure.

The empirical analysis identified substantial differences in hard-drive encryption, authentication-related practices, password complexity, and several endpoint and security-control measures. These findings support the inclusion of technical protection as a distinct framework dimension. However, controls such as VPN deployment, website allowlisting, and specific EDI protocols are framework recommendations rather than direct findings from the present survey and should therefore be validated according to organizational requirements.

\subsection{Physical and Environmental Security}

The Physical and Environmental Security dimension protects the infrastructure on which EDI systems and supporting information assets depend. Logical security controls cannot fully protect systems if unauthorized individuals can physically access servers, network equipment, endpoints, or sensitive records.

Recommended controls include restricted physical access, visitor management, surveillance, equipment protection, secure disposal of storage media, environmental safeguards, and appropriate protection of workstations and network infrastructure. Physical controls should be aligned with asset criticality and the operational environment.

This dimension complements the technical and governance dimensions by addressing risks that originate outside the logical EDI environment. It is particularly relevant to organizations operating their own infrastructure or maintaining sensitive EDI processing and networking equipment.

\subsection{Continuous Monitoring and Improvement}

Continuous Monitoring and Improvement forms the feedback mechanism of the EDI-RMF. Rather than treating security implementation as a one-time activity, organizations should continuously collect evidence regarding control effectiveness, security events, incidents, vulnerabilities, compliance status, and changes in the threat environment.

This dimension includes security-log monitoring, periodic control assessments, vulnerability assessment, incident detection and response, business continuity testing, recovery exercises, internal audits, and periodic policy review. Findings from these activities should be fed back into risk assessment and governance decisions.

The empirical relationships between SCP, TP, and ORTA further support this feedback-oriented approach. Security preparedness and perceived threats were significantly associated with organizational risk-taking ability. Accordingly, the framework treats risk perception and security-control effectiveness as inputs to ongoing risk decisions rather than static attributes of an organization.

\subsection{Risk Assessment and Feedback Cycle}

The five dimensions of the EDI-RMF operate within a continuous risk-management cycle. Organizations should first identify EDI assets, information flows, business dependencies, threats, vulnerabilities, and potential impacts. The identified risks are then evaluated against organizational risk criteria to prioritize appropriate controls.

Control implementation is followed by monitoring and assessment. Evidence from security logs, audits, vulnerability assessments, incidents, continuity exercises, and user-awareness activities is used to determine whether controls remain effective. Identified deficiencies are prioritized for remediation, while significant changes in technology, business processes, trading partners, or regulatory requirements trigger reassessment.

The resulting feedback is returned to the governance layer, where policies, responsibilities, risk priorities, and resource allocation can be revised. This creates a closed-loop process:
$
Identify Risk \rightarrow Assess Risk \rightarrow
Implement Controls \rightarrow Monitor \rightarrow Evaluate
\rightarrow Improve \rightarrow Reassess
$

This cycle allows the EDI-RMF to accommodate changing threats and organizational conditions without requiring a complete redesign of the security architecture.

\subsection{Application Across EDI-Enabled Sectors}

The EDI-RMF is intended as a sector-adaptable framework rather than a healthcare-specific security model. The relative priority of individual controls may vary according to the sensitivity of exchanged information, regulatory obligations, transaction volume, business dependencies, technical architecture, and organizational risk tolerance.

For example, healthcare environments may require stronger emphasis on protection of sensitive health information and accountability, while financial, manufacturing, and logistics environments may place greater emphasis on transaction integrity, availability, supply-chain continuity, and partner connectivity. The same five framework dimensions can therefore be retained while control selection and implementation priorities are adapted to the specific risk context.

\subsection{Framework Contribution}

The principal contribution of the EDI-RMF is the integration of organizational and technical security considerations into a single risk-management structure informed by empirical evidence. The framework does not assume that a particular professional background determines security maturity. Instead, it identifies organizational capabilities that can be assessed, implemented, monitored, and improved across EDI-enabled environments.

The framework consequently provides a bridge between the empirical findings and practical EDI security management. The observed differences in security-control preparedness, organizational practices, and threat perception inform the selection and prioritization of framework dimensions, while the continuous feedback cycle provides a mechanism for translating assessment results into ongoing security improvement.

\section{Conclusion}

This study assessed information-security preparedness in Electronic Data Interchange (EDI)-enabled environments by comparing General IT Professionals (GIT) and Healthcare EDI Professionals (HEP) in Nepal. HEP respondents reported significantly higher Security Control Preparedness (SCP) and Organizational Risk-Taking Ability (ORTA), whereas GIT respondents reported higher Threat Perception (TP). Significant differences were also observed in several practices, including risk assessment, business continuity, incident response, audit logging, encryption, removable-media controls, and authentication.

SCP was positively associated with ORTA, while TP was negatively associated with it. A regression model incorporating both constructs explained 41.6\% of the observed variance in ORTA. These associations do not establish causality but highlight the importance of considering both implemented security controls and organizational risk perceptions.

Based on these findings, the study proposes the EDI Risk Management Framework (EDI-RMF), integrating governance and policy, human and organizational security, technical safeguards, physical protection, and continuous monitoring and improvement. The framework translates observed security gaps into a structured approach to risk assessment and control prioritization. Although further validation is required, it provides a foundation for strengthening information-security risk management across EDI-enabled sectors.

\section{Limitations and Future Work}

This study has several limitations. Its cross-sectional design and purposive sample of 100 respondents limit causal interpretation and generalizability beyond the surveyed groups. Furthermore, the findings reflect self-reported practices rather than independently verified security effectiveness. The healthcare group represents a comparative context, not a universal proxy for security expertise. The study also does not comprehensively assess third-party EDI dependencies, and the proposed EDI-RMF has not yet been validated through organizational implementation.

Future research should use larger, multi-sector and multi-country samples, longitudinal designs, and independent technical assessments to validate reported security practices. Further investigation of the negative association between TP and ORTA is also warranted. Pilot implementations of the EDI-RMF should evaluate its usability and effectiveness using measurable indicators, including control coverage, risk reduction, incident-response performance, and audit outcomes.

\end{document}